\documentclass[preprint,showpacs,preprintnumbers,amsmath,amssymb]{revtex4}
\usepackage{graphicx,color}
\usepackage{epsfig}
\def\be{\begin{eqnarray}}
\def\ed{\end{eqnarray}}
\def\non{\nonumber}
\def\ga{\gamma}
\def\e{\epsilon}
\def\ve{\varepsilon}

\begin{document}

\title{\Large \bf  Forward-backward asymmetry of top quark\\
in diquark models }

\date{\today}

\author{ \bf Abdesslam Arhrib$^{1,2}$\footnote{Email:aarhrib@ictp.it},
Rachid Benbrik$^{2,3,4}$\footnote{Email:rbenbrik@mail.ncku.edu.tw} and
Chuan-Hung Chen$^{3,4}$\footnote{Email:physchen@mail.ncku.edu.tw} }

\affiliation{$^1$ D\'epatement de Math\'ematique, Facult\'e des
Sciences and Techniques, Universit\'e Abdelmalek Essa\^adi, B. 416,
Tangier, Morocco \\
$^2$ LPHEA, Facult\'e des
Sciences-Semlalia, B.P. 2390 Marrakesh, Morocco\\
$^{3}$ Department of Physics, National Cheng-Kung University, Tainan
701, Taiwan \\
$^{4}$National Center for Theoretical Sciences,  Hsinchu 300, Taiwan}

\begin{abstract}
Motivated by the recent unexpected large forward-backward asymmetry
(FBA) of the top quark observed by D$\O$ and CDF at the Tevatron, we
investigate a possible explanation for the anomaly within the
framework of diquark models. In the diquark models,
the top-quark pair production is mediated by the u-channel diagram.
It is found that the color-triplet diquark can generate the FBA of 20\%
when the constraint from the cross section of the top-quark pair
production is taken into account.
\end{abstract}

\maketitle

Although most experimental data are consistent with the Standard Model (SM)
predictions, it is believed that the SM is just
an effective theory of a more fundamental one yet to be discovered.
For a correct understanding of the hierarchy problem of the Higgs mass
and Planck scale, neutrinos masses, matter-antimatter asymmetry,
dark matter, etc, new physics beyond the SM should be included.
Theorists and experimentalists have put a lot of effort to
investigate the existence of new physics. If such new physics exists,
it can be probed either directly at collider or indirectly through
precise measurements. Intriguingly, the new measurements at
 Tevatron on the  forward-backward asymmetry (FBA) in
 the top-quark pair production at $\sqrt{s}=1.96$ TeV may provide the clue
for the existence of  new physics.

When D$\O$ Collaboration published the first measurement on the FBA
in top-quark pair production in the $p\bar p$ laboratory frame with
0.9 fb$^{-1}$ of data, an unexpected large FBA of top-quark was
found \cite{D0_PRL100}. Subsequently, CDF also observed the same
phenomenon by using 1.9 fb$^{-1}$ \cite{CDF_PRL101}.
Furthermore, with an integrated luminosity of $3.2$ fb$^{-1}$,
the updated CDF's result in the $p\bar p$ laboratory frame is given
by \cite{CDFnote}
 \be
 A^{t}_{FB}=0.193 \pm 0.065(\rm stat) \pm 0.024 (\rm syst)\,.
 \ed
The indication of large FBA of top-quark is not smeared by
statistic. In the SM, since the top-quark pair production is
dominated by the strong interaction QCD contribution, due to
C-parity invariance, a vanishing FBA at the leading order (LO) is
expected. However, a nonvanishing FBA can be induced at the
next-to-leading order (NLO) \cite{AKR} and is given by $A^{t}_{\rm
FB}=0.05 \pm 0.015$. Comparing the SM prediction with the latest CDF's
measurement, although the deviation of data from the SM result is less
than $3\sigma$, the difference could actually originate from new
physics contributions. Motivated by the recent measurements of D$\O$
and CDF at Tevatron, several possible explanations to the observed
FBA have been proposed in several
studies.~\cite{FBA1,FBA2,FBA3,FBA4,FBA5,FBA6,Dorsner:2009mq}.
%
%
In this paper, we investigate the impact of colored scalars
``diquarks'' on the FBA of the $t\bar{t}$ production at Tevatron,
where the diquark will  contribute to the $t\bar{t}$ production
through a u-channel diagram.

It is known that  the scalar sector of the SM has not been tested
directly by any experiments, it is plausible to assume the existence of
other possible scalars within $SU(3)_{c} \times
SU(2)_{L} \times U(1)_Y$ gauge symmetry.
Theoretically, exotic color states are well motivated.
For example, one can find: color non-singlet scalar
fields for CP violation \cite{Barr_PRD34,BF_PRD41}, color scalar
quarks in supersymmetry with R-parity violation
\cite{Barbier:2004ez}, supersymmetric $SU(2)_L\times SU(2)_R\times
SU(4)_c$ model that embeds the seesaw mechanism for neutrino masses
where a color sextet Higgs fields appears at TeV range \cite{moha}.
Consequently, the general scalar representations could be ${\bf( 1,
2)}_{1/2}$, ${\bf (8, 2)}_{1/2}$, ${\bf (6, 3)}_{1/3}$, ${\bf (6,
1)}_{4/3,1/3,-2/3}$, ${\bf(3, 3)}_{-1/3}$, ${\bf(3,
1)}_{2/3,-1/3,-4/3}$, where the first (second) argument in the
brackets denotes the representation in color (weak isospin) space
and the number in the subscript corresponds to the hypercharge of
$U(1)_Y$ \cite{Barr_PRD34,BF_PRD41,MW}. Besides the SM Higgs
doublet, it has been shown in Ref.~\cite{MW} that when the
hypothesis of minimal flavor violation (MFV) is imposed and if
the scalar is flavor singlet \cite {Arnold:2009ay}, only the
representation ${\bf (8, 2)}_{1/2}$ could avoid FCNCs at the tree level.
As a result, the couplings of color octet to quarks are associated
with the quark masses. Therefore, the s-channel induced process
$u\bar u (d\bar d)\to t\bar t$ in the color octet model is
negligible. In addition, due to the suppression of
Cabibbo-Kobayashi-Maskawa (CKM) matrix element $V_{td}$, the
t-channel $d\bar d\to t\bar t$ process induced by charged
color-octet turns out to be rather small. In the following analysis,
we will concentrate on the situation of color triplet and sextet.

\begin{figure}[bpth]
\includegraphics*[width=4.5 in]{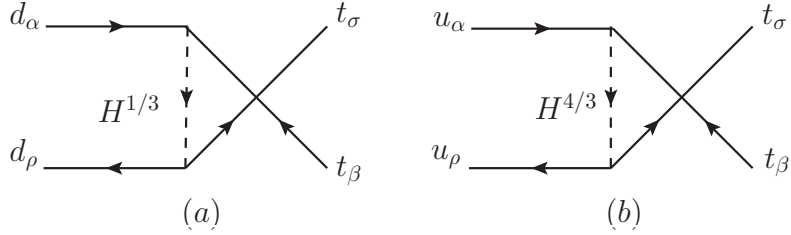}
\caption{Feynman diagrams for (a) $d\bar d\to t\bar t$ generated
by models (1)-(5) and (b) $u\bar u \to t\bar t$ generated by models
(6) and (7).}
 \label{fig:ut}
 \end{figure}

For understanding what kind of colored scalars could contribute to
FBA of top quark, we display various possible scalar diquarks in
Table~\ref{tab:diquark} \cite{Chen09}, where the second column in
the table denotes the representations of the diquarks under
$SU(3)_c\times SU(2)_L \times U(1)_Y$, the third column gives the
interactions of quarks and diquarks and the fourth column displays
the relation of couplings in flavors. The notations used in the Table are as follows: the indices $(i, j)$ and $(\alpha, \beta)$ stand for the flavors and colors, $Q$ and $u(d)$ denote the $SU(2)_L$ doublet and singlet quarks,  $\epsilon^{\alpha\beta\gamma} H_\gamma$ and $H^{\alpha \beta}$ represent the color triplet and sextet, respectively,  $f^c= C \bar f^T$ with $C$ being the charge conjugation operator and $\epsilon =i \tau_2$. The involved couplings are
 considered as free parameters.
\begin{table}[hptb]
\caption{Various diquark models for the $d\bar d \to t\bar t$ and $u\bar u \to t\bar t$.}
\label{tab:diquark}
\begin{ruledtabular}
\begin{tabular}{cccc}
Model & H  & Interaction & flavor symmetry
 \\ \hline
(1) & $(3,1,-1/3)$ & $f_{ij} \bar Q^c_{i\alpha} \e Q_{j\beta}\ve^{\alpha\beta\ga} H_{\ga}$ & $f_{ij}=f_{ji}$  \\
(2) & $(3,1,-1/3)$ & $f_{ij} \bar d_{i\alpha} u^c_{j\beta}\ve^{\alpha\beta\ga} H^\dagger_{\ga}$ & $--$ \\
(3) & $(6,1,1/3)$ & $f_{ij} \bar Q^c_{i\alpha} \e Q_{j\beta} H^{\dagger\alpha\beta}$ & $f_{ij}=-f_{ji}$  \\
(4) & $(6,1,1/3)$ & $f_{ij} \bar d_{i\alpha} u^c_{j\beta} H^{\alpha\beta}$ & $--$  \\
(5) & $(3,3,-1/3)$ & $f_{ij} \bar Q^c_{i\alpha} \e H_{\ga} Q_{j\beta}\ve^{\alpha\beta\ga} $ & $f_{ij}=-f_{ji}$  \\
(6) & $(3,1,-4/3)$ & $f_{ij} \bar u_{i\alpha}  u^c_{j\beta}\ve^{\alpha\beta\ga} H^{\dagger}_{\ga}$ & $f_{ij}=-f_{ji}$  \\
(7) & $(6,1,4/3)$ & $f_{ij} \bar u_{i\alpha}  u^c_{j\beta} H^{\alpha\beta}$ & $f_{ij}=f_{ji}$  \\
\end{tabular}
\end{ruledtabular}
\end{table}
 From the table, we find that the $t\bar t$ pair in
$p\bar p$ collisions can be produced via
$d\bar d\to t\bar t$ and $u\bar u \to t\bar t$ processes, where the
former is represented by  models (1)-(5) while the latter is
dictated by models (6) and (7). To be more specific, we sketch the associated
Feynman diagrams in Fig.~\ref{fig:ut}. Owing to the
diquarks being color triplet and sextet in each diagram, for not
showing the specific diquark, we have suppressed their color
indices. In addition, one can also easily find that the couplings of
diquarks to quarks are chiral, in other words,  only left- or
right-handed quarks are involved in the diquark models. Due to the
chiral interactions, it is expected that a large FBA in $t\bar t$
production could be induced by the diquark contribution.

Since our purpose is to illustrate the impact of diquarks on the FBA of
top quark production, we will not demonstrate the results of all models shown
in Table~\ref{tab:diquark}. Instead, we select the diquarks of
$(3,1,-4/3)$ and $(6,1,4/3)$ as representative, where the $t\bar t$
pair is produced through a u-channel diagram with diquark exchange as
depicted in  Fig.~\ref{fig:ut}.
Accordingly, the interactions of color-triplet and sextet diquarks
with quarks are respectively written by
 \be
 {\cal L}_{\bf 3}&=&-f^{\bf 3}_{ij} \bar u_{i\alpha} P_L u^c_{j\beta}
 \epsilon^{\alpha\beta\gamma} H^\dagger_{\bf 3\gamma} + h.c. \,, \non\\
 {\cal L}_{\bf 6}&=&-f^{\bf 6}_{ij} \bar u_{i\alpha} P_L u^c_{j\beta}
  H^{\alpha\beta}_{\bf 6} + h.c.
 \label{eq:lang}
 \ed
where the indices $i$ and $j$ are the quark flavors, $f^{\bf
3}_{ij}=-f^{\bf 3}_{ji}$, $f^{\bf 6}_{ij}=f^{\bf 6}_{ji}$,
$(\alpha,\beta,\gamma)$ stand for the color indices,
$\epsilon^{\alpha\beta\gamma}$ is antisymmetric tensor in color
space and $P_{L(R)}=(1\mp \gamma_5)/2$ is the helicity projection.
In terms of Eq.~(\ref{eq:lang}), the process $u\bar u \to t\bar
t$ could proceed through the following interactions
 \be
 {\cal L}&=& 2f^{\bf 3}_{ut} \bar u_\alpha P_L t^c_\beta
 \epsilon^{\alpha\beta\gamma}H^\dagger_{\bf 3\gamma} +
 2f^{\bf 6}_{ut} \bar u_\alpha P_L t^c_\beta H^{\alpha\beta}_{\bf 6} + h.c.
 \label{eq:ut_int}
 \ed

Before presenting the amplitude of $u\bar u\to t\bar t$ mediated by
gluon and the diquark $H$, let us first define a convenient notation
for the following calculations.  The momenta of the incoming
up-quark and anti-up-quark, outgoing top and outgoing anti-top
quarks are denoted by $p_u$, $p_{\bar{u}}$, $p_t$ and $p_{\bar{t}}$
respectively such that $p_u+p_{\bar{u}}=p_t+p_{\bar{t}}$. The
momentum
 can be written as:
\begin{eqnarray}
&&p_{u,\bar{u}}= \frac{\sqrt{\hat s} }{2}(1,0,0,\pm 1)\nonumber\\
&&p_{t,\bar{t}}= \frac{\sqrt{\hat s} }{2}(1,\pm\beta \sin\theta,0,\pm
\beta\cos\theta)
\label{eq:mom}
\end{eqnarray}
where $\beta^2 = 1-4m_t^2/\hat s$ and $\theta$ is
the scattering angle in the center-of-mass frame of the partons.
Neglecting the up-quark masses of the incoming partons,
the Mandelstam variables are defined as follow
\begin{eqnarray}
\hat s &=&  (p_u+p_{\bar{u}})^2 = (p_t+p_{\bar{t}})^2\,,  \nonumber \\
\hat t &=& (p_u-p_t)^2 = (p_{\bar{u}}- p_{\bar{t}})^2
= m_t^2 - \frac{\hat s}{2} \left( 1 - \beta \cos\theta
                                              \right )\,,
\nonumber \\
\hat u &=& (p_u-p_{\bar{t}})^2 = (p_{\bar{u}}-p_t)^2
=m_t^2 - \frac{\hat s}{2} \left( 1 + \beta \cos\theta
                                              \right )\,.
\label{eq:mandel}
\end{eqnarray}

By combining both the gluon SM contribution  and the diquark
contribution at leading order, the amplitude can be written as
 \be
{\cal M}_{SM+H}(u\bar u\to t\bar t)&=& \frac{g^2_s}{\hat{s}}\left[ \bar
v_{u} (p_{\bar u}) \ga^{\mu} T^a u_{u}(p_u) \right]\,\left[ \bar
u_{t}(p_t) \ga_{\mu} T^a v_{t}(p_{\bar t}) \right]\non
\\
&-& \frac{|2f_{ut}|^2}{2(\hat{u}-m^2_H)}\left[\bar v_{u}(p_{\bar
u}) \ga_\mu P_R u(p_u) ~ \bar u_{t}(p_t) \ga^\mu P_R v_t(p_{\bar t})
\right. \non \\
&&\left.+ \epsilon_{c} \bar u_{t\alpha}(p_t)\ga^\mu P_R v^\beta_t(p_{\bar t})~ \bar
v_{u\beta}(p_{\bar u}) \ga_\mu P_R u^{\alpha}_u(p_u) \right]\,,
  \ed
where $\epsilon_c=-(+)$ denotes the color triplet (sextet), $T^a$
are the generators of $SU(3)_c$ group, $g_s$ is the gauge
coupling constant. In addition, we have adopted the Fierz transformation $\bar
q_1 P_L q_2 \bar q_3 P_R q_4 =(\xi/2) \bar q_1 \ga^\mu P_R q_4 \bar q_3
\ga_\mu P_L q_2$ and $\bar q^c_1 \ga_\mu P_L q^c_2=\xi \bar q_2 \ga_\mu P_R
q_1$ in which  $\{q_i\}$ are the spinors and $\xi=+(-)$ when $\{q_i\}$ are
regarded as c-numbers (field operators). Then, the squared amplitude takes the
following form
 \be
\sum \overline{ \left| {\cal M}_{SM+H} \right |^2 } &=& \frac{4\pi^2
\alpha_s^2}{N^2_c} (N^2_c-1) \left(1+\beta^2
\cos^2\theta + \frac{4m^2_t}{\hat{s}} \right) \non \\
&-& \frac{\pi \alpha_s}{N^2_c}\epsilon_c \left( \frac{N^2_c-1}{2}\right)
\frac{f^{\prime^ 2}_{ut}\hat{s} }{\hat{u}-m^2_H} \left((1+\beta\cos\theta)^2
+ \frac{4m^2_t}{\hat{s}} \right) \non \\
&+& \frac{1}{8N^2_c}N_{c} (N_c + \epsilon_c) \frac{f^{\prime^ 4}_{ut}
\hat{s}^2}{(\hat{u} -m^2_H)^2}  (1+\beta\cos\theta)^2
 \label{eq:amp2}
 \ed
where we have already summed over final state color and averaged
over the initial spin and color, $f'_{ut}=2f_{ut}$ and $N_c=3$. We note
that since the propagator in the $u$-channel diagram depends on the
scattering angle $\theta$,
the FBA may arise not only from $\cos\theta$ term in
Eq.~(\ref{eq:amp2}), but also from the constant term and the
$\cos^2\theta$ term. The partonic differential cross section for the
subprocess $u\bar u\to t\bar t$ in the parton rest frame is given by
 \be
 \frac{d\hat\sigma}{d\cos\theta} &=&
\frac{\beta}{32\pi \hat s} \sum\overline{ \left| {\cal M}_{SM+H} \right
|^2 } \label{eq:dcross}
 \ed
To obtain the hadronic cross section $\sigma$, the partonic one is
then convoluted with the parton distribution functions. In our study
we have used PDF distribution CTEQ6 \cite{pdf} at the leading order
and set the renormalization and factorization scales to
$\mu_F=\mu_R=m_t$. Accordingly, the FBA in top-quark pair
production is defined by
\be
A^{t}_{FB} = \frac{\sigma(\eta \ge 0)-\sigma(\eta\le 0) }{
\sigma(\eta \ge 0) + \sigma(\eta\le 0) }
\label{eq:fba}
 \ed
 where $\eta$ denotes the rapidity
of top quarks
 and is defined by $\eta= 2 \tanh^{-1}(\beta
 \cos\theta)$.


Before presenting our numerical analysis, it is worth discussing the strict
constraints by other low energy processes such as $D^0-\bar D^0$
mixing. It has been studied in Ref.~\cite{Chen09} that the diquark
of  (6,1,4/3) representation could contribute to
$D^0-\bar D^0$ mixing by a tree-diagram while  the diquark of
(3,1,-4/3) representation would contribute through the box-diagram.
Nevertheless, since the related parameter for the former is $f_{uc}$
while the latter is $f_{ut}f^*_{ct}$, by setting $f_{ct}$ to be
small enough, $f_{ut}$ could escape the direct constraint from $D^0-\bar
D^0$ mixing. In other words, $f_{ut}$ could be taken as a free
parameter and could be of the order of unity in our study.
The only involved free parameters are the coupling $f_{ut}$ and
the diquark mass $m_H$, which could be constrained with the use of
the experimental measurement of $t\bar t$ cross section
$\sigma(p\bar p \to t\bar t)$ from
\cite{Abazov:2009ae}
\begin{equation}
 \sigma (p\bar p \to t \bar t)^{\rm exp} = 7.5\pm 0.31\;({\rm stat})\; \pm 0.34\;({\rm
   syst}) \;\pm 0.15 \; ({\rm th}) \;  \ \ {\rm pb}\,,
\label{eq:data}
\end{equation}
where the SM prediction is $\sigma(p\bar p\to t\bar t)^{\rm
SM}=6.73^{+0.71}_{-0.79}$ pb \cite{ttnlo}. In our numerical
estimation, 
we use $m_{t} = 172.5$ GeV with $\alpha_s = 0.1095$
we take as input the experimental value
given by Eq.~(\ref{eq:data}) with $1\sigma$ errors.
In addition, for matching the cross section of $p\bar
p\to t\bar t$ up to the next-to-leading order in the SM \cite{ttnlo}, a 
QCD K-factor of 1.3 has been multiplied everywhere in our calculations.

\begin{figure}[phtb]
\includegraphics*[width=4. in]{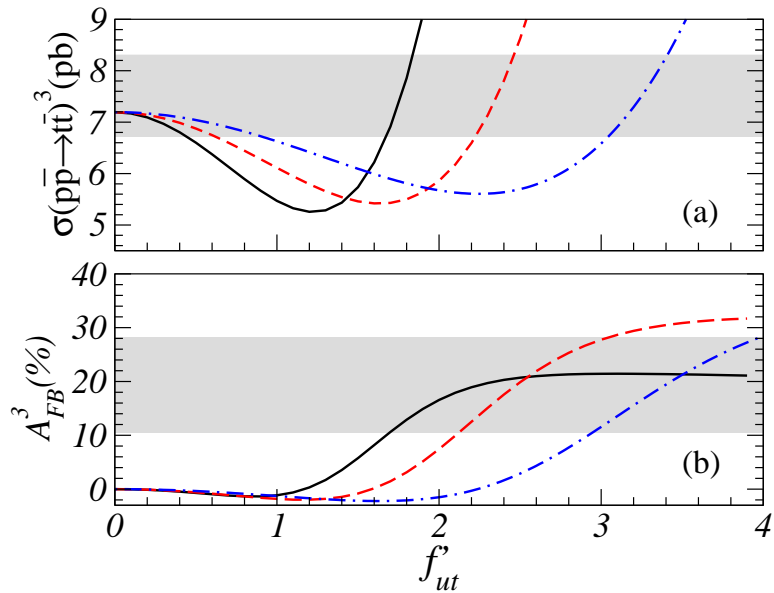}
\caption{(a) Cross section for inclusive top-quark pair production
and (b) FBA of top-quark in triplet diquark model, where the solid,
dashed and dash-doted lines denote $m_H=300$, 500 and 800 GeV,
respectively. The bands in the plots denote the experimental data
with $1\sigma$ errors. }
\label{fig:triplet}
\end{figure}

With the above inputs, we are now ready to present our
numerical analysis for the
contributions of  color-triplet and -sextet diquarks separately. In
terms of Eqs.~(\ref{eq:amp2}) and (\ref{eq:dcross}) and with
$\epsilon_c=-1$, the inclusive cross section for $t\bar t$
production mediated by gluons and triplet diquark is displayed in
Fig.~\ref{fig:triplet}(a), where the solid, dashed and dash-dotted
lines denote the influence of triplet diquark with $m_H=300$, 500
and 800 GeV, respectively. The band shown in the plot is the
experimental measurement of $t\bar t$ cross section
 with $1\sigma$ errors. Clearly, $\sigma(p\bar p\to t\bar t)$ is
sensitive to the diquark contribution. Moreover, by
Eq.~(\ref{eq:fba}), we present the FBA in top-quark pair production in
Fig.~\ref{fig:triplet}(b).
When we contrast Fig.~\ref{fig:triplet}(a) with
Fig.~\ref{fig:triplet}(b), it is clear that with the same values of
parameters, triplet diquark can fit both the FBA of $(10-30)\%$
 as well as $t\bar t$ cross section  $\sigma(p\bar p\to t\bar t)^{\rm exp}$
simultaneously. As it can be seen from  Fig.~\ref{fig:triplet}, this could
happen when $f'_{ut}$ is located in the range $(1.67,\, 1.83)$,
$(2.22,\, 2.43)$ and $(3.04,\, 3.39)$
for $m_H=300$, 500 and 800 GeV, respectively.

\begin{figure}[hptb]
\includegraphics*[width=4. in]{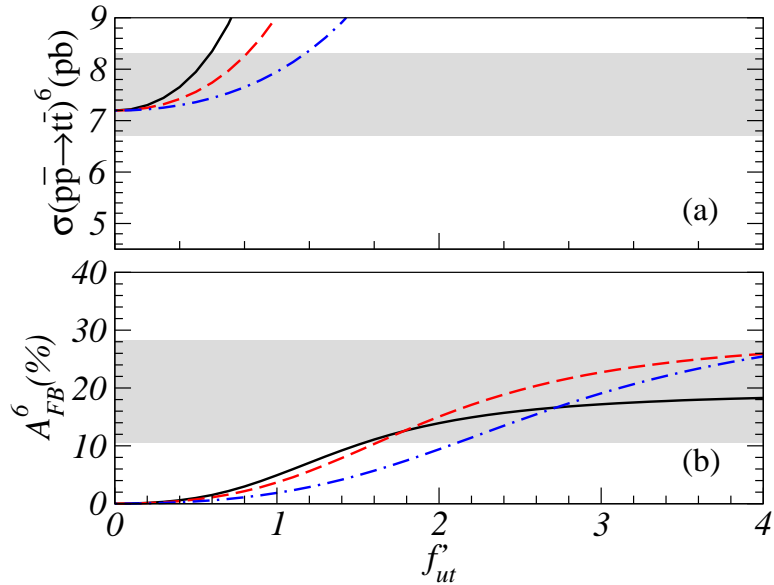}
\caption{Legend is the same as Fig.~\ref{fig:triplet} but for sextet
diquark model.}
 \label{fig:sextet}
\end{figure}

Similarly, with $\epsilon_{c}=+1$, the $t\bar t$ cross section
$\sigma(p\bar p\to t\bar t)$
 mediated by sextet diquark as a function of $f'_{ut}$ is illustrated in
Fig.~\ref{fig:sextet}(a), in which the solid, dashed and dash-dotted
lines represent the contributions from $m_H=300$, 500 and 800 GeV,
respectively. In Fig.~\ref{fig:triplet}(a) and
Fig.~\ref{fig:sextet}(a), we can see that the behavior of
triplet and sextet contribution in $t\bar t$ production is quite different.
In some region of $f'_{ut}$, the interference between triplet diquark
and SM-gluon contribution is destructive. In contract,
the interference between the sextet diquark and SM-gluon contribution
 is all the time constructive.
This different behavior between triplet and sextet is mainly due to
the sign of the  2nd term in Eq. (\ref{eq:amp2})
for which the triplet diquark is negative
while the sextet diquark is positive.
Furthermore, we show the associated
$A^{t}_{FB}$ defined by Eq.~(\ref{eq:fba}) in
Fig.~\ref{fig:sextet}(b). From Figs.~\ref{fig:sextet}(a) and
\ref{fig:sextet}(b), one can easily see that the sextet diquark
could not fit the current data of $A^{t}_{FB}$ and $\sigma(p\bar p \to
t\bar t)^{\rm exp}$ simultaneously.

\begin{figure}[hptb]
\includegraphics*[width=5.5 in]{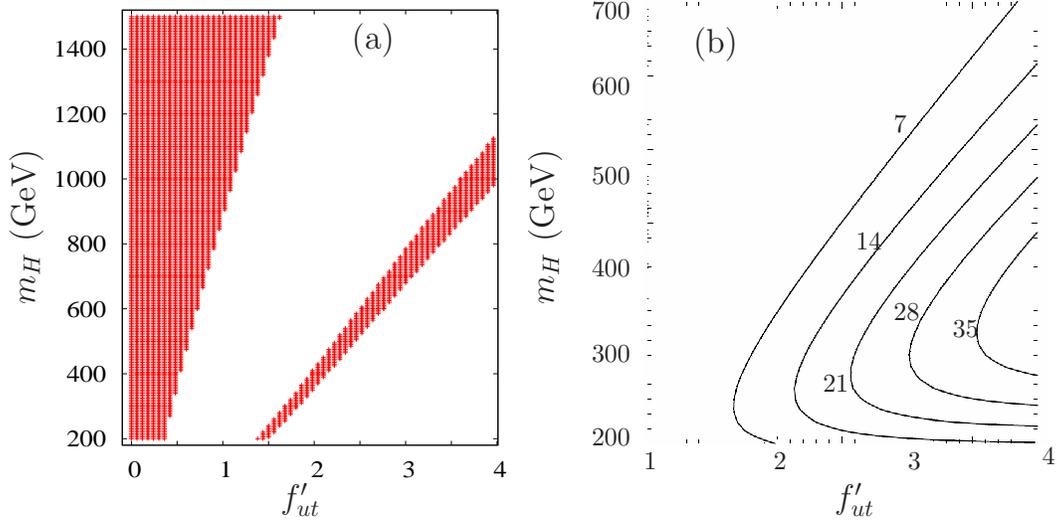}
\caption{(a) The allowed range of $f'_{ut}$ and $m_H$ by fitting the measurement of
$\sigma(p\bar p\to t\bar t)^{\rm exp}$ with $1\sigma$ errors. (b)
Contour for FBA as a function of $f'_{ut}$ and $m_H$, where the results in (a) have been applied.
}
\label{fig:contour}
\end{figure}

According to previous analysis, we have shown that color-triplet
diquark plays an important role on the top-quark FBA. To be more
clear, we have shown how $\sigma(p\bar p\to t\bar t)$ and $A^t_{FB}$
are affected by the free parameters of the diquark models.
For this purpose, we perform a systematic scan over
the parameters $f^{\prime}_{ut}$ and $m_H$ of the triplet diquark. In
Fig.~\ref{fig:contour}(a), we show the allowed values of $f'_{ut}$
and $m_H$ that satisfy the measurement of
$\sigma(p\bar p\to t\bar t)$ within $1\sigma$ errors.
Furthermore, with those allowed values,
we make a two-dimensional contour as a function of $f'_{ut}$ and
$m_{H}$ in Fig.~\ref{fig:contour}(b), where the labels in the plot
denote the corresponding values of $A^{t}_{FB}$.
From the contour plot, not only can one
easily see the influence of triplet diquark on the FBA of top-quark,
but also one can understand the correlation between $f'_{ut}$ and $m_{H}$.

\begin{figure}[hptb]
\includegraphics*[width=4.8 in]{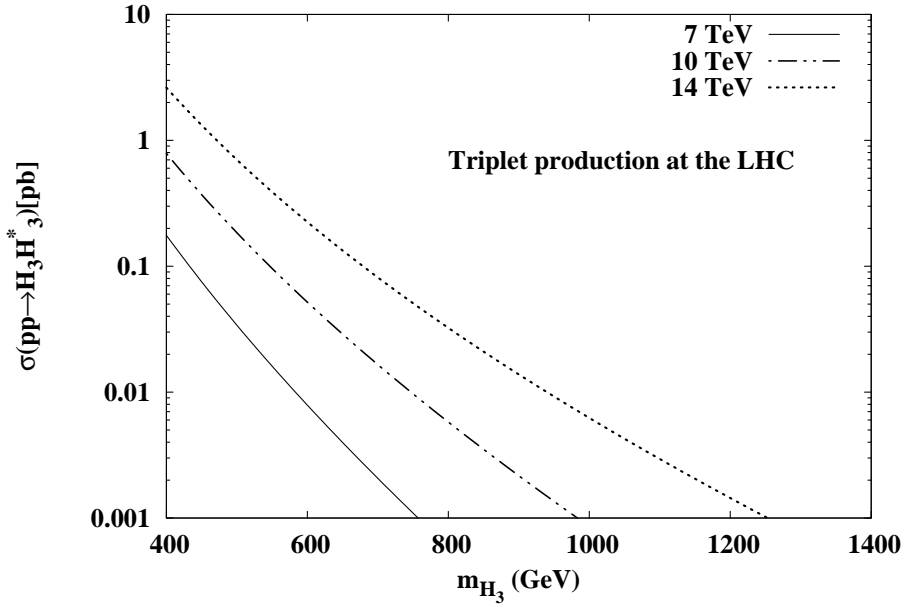}
\caption{Hadronic production cross section of triplet $H_{\bf 3}$ as a 
function of triplet mass $m_{H_{\bf 3}}$ at the LHC 
for $\sqrt{s} = 7, 10$ and $14$ TeV.}
 \label{fig:production}
\end{figure}

In the present analysis, we have shown the effects of colored scalars
on the FBA in top-quark production. If the diquarks particles with a mass
of the order $O(100\ \rm GeV)$ exit in the universe,
they could be produced at hadron
colliders such as Tevatron and LHC. Those diquarks can be produced
 through the annihilation of a pair of up quarks $uu\to H_{\bf6}$
\cite{MOY_PRD77,diqprod} and in association with
 top-quark through $u g\to \bar t H_{\bf 3, 6}$
\cite{FBA6}. 
Scalar diquark can also be produced in pair, 
such production process proceeds through the QCD couplings of the diquark. 
The general cross section of scalar diquark pair production is 
given in the first reference of~\cite{diqprod} at the tree-level. 
Here, we plot in Fig.~\ref{fig:production}, scalar triplet pair 
production at the LHC with different energies. The factorization 
scale has been set to $\mu_{F} = m_{H_{\bf 3}}$. As we can see, the total 
production cross section of the triplet scalar can reach 17fb, 78fb 
and 2.6pb for 7, 10  and 14 TeV, respectively for Diquark mass of 400 GeV. 
Note that for LHC with 14 TeV with low luminosity of 1$fb^{-1}$ 
would lead to 2600 raw events which could be enough to extract a signal.

If kinematically allowed, 
the scalar diquark would then decay into $tt\bar{t}\bar{t}$ 
in the case of pair production which would lead to same-sign dileptons signal. 
Single production of scalar diquark can lead to double top
$tt$ and single top $tu$, $tc$ plus jet production depending on its
mass, coupling and the representation of colored scalar. Note that
some decay channel of the diquark such as $tt$, which could lead to
same-sign dileptons plus multijets if both top-quarks decay
semileptonically, does not exist in SM and would be, in principle,
easily distinguished from the SM background. At the Tevatron, the
production rate of diquark is rather small, whereas the production
rate of diquark is rather large at the LHC \cite{diqprod}. Hence, the LHC
provides a good environment to discover the diquarks and study their
properties.
In summary, an unexpected large FBA of top-quark has been observed
by CDF and D{\O} Collaborations \cite{D0_PRL100,CDF_PRL101,CDFnote}.
To account for this anomaly, we study the contribution of diquarks
to the FBA in top-quark production.
We find that the parameters of diquark models
involved in $u\bar u \to t\bar t$ can escape the strict limit
imposed by $D^0-\bar D^0$ mixing, therefore the measured cross section of
top-pair production at the Tevatron is the main constraint. We show
that with the current data of $\sigma(p\bar p\to t\bar t)$ within $1\sigma$
 errors, only
color triplet diquark can give a large and consistent FBA with the
updated measurement of CDF \cite{CDFnote}, while the color-sextet
representation can not fit  simultaneously the $A^t_{FB}$ and
$\sigma(p\bar p\to t\bar t)^{\rm exp}$. \\

\noindent {\bf Note added:} While we were finishing the present
work, we received a paper \cite{Dorsner:2009mq} 
dealing with similar subject.
Our result agrees with \cite{Dorsner:2009mq} in the case of color-triplet
diquark. However,
our results for triplet and sextet diquarks are consistent with
the paper \cite{Jung:2009pi}.\\

\noindent{\bf Acknowledgments}

 We thank Dr. J. Shu and Dr. J.~F.~Kamenik for their useful discussions on the color factor and the sign convention in the interference between SM and diquark contributions.
A.A thanks LPTA University Montpellier-II
for hospitality extended to him during his
visit, project AI $N^0$ MA-186-08,
where part of this work has been done.
R.B is supported by National Cheng
Kung University Grant No. HUA 97-03-02-063. R.B acknowledges the KEK
theory exchange program for physicists in Taiwan and the very kind
hospitality at KEK. C.C.H is supported by the National Science
Council of R.O.C under Grant \#s: NSC-97-2112-M-006-001-MY3.


\end{document}